# Developing Classification Indices for Chinese Pulse Diagnosis

## JIAN-JUN SHU[*] and YUGUANG SUN

School of Mechanical & Aerospace Engineering,

Nanyang Technological University, 50 Nanyang Avenue, Singapore 639798

## Summary

*Aim*: To develop classification criteria for Chinese pulse diagnosis and to objectify the ancient diagnostic technique.

*Methods*: Chinese pulse curves are treated as wave signals. Multidimensional variable analysis is performed to provide the best curve fit between the recorded Chinese pulse waveforms and the collective Gamma density functions.

*Results*: Chinese pulses can be recognized quantitatively by the newly-developed four classification indices, *that is*, the wave length, the relative phase difference, the rate parameter, and the peak ratio. The new quantitative classification not only reduces the dependency of pulse diagnosis on Chinese physician's experience, but also is able to interpret pathological wrist-pulse waveforms more precisely.

*Conclusions*: Traditionally, Chinese physicians use fingertips to feel the wrist-pulses of patients in order to determine their health conditions. The qualitative theory of the Chinese pulse diagnosis is based on the experience of Chinese physicians for thousands of years. However, there are no quantitative theories to relate these different wrist-pulse waveforms to the health conditions of patients. In this paper, new quantified classification indices have been introduced to interpret the Chinese pulse waveform patterns objectively.

## Introduction

Chinese pulse diagnosis in Traditional Chinese Medicine (TCM) has been practiced for more than 2000 years[1,2]. Chinese physicians use fingertips to feel the wrist-pulses of patients in order to determine their health conditions. The wrist-pulse has been recognized as the most fundamental signal of life, containing vital information of health activities.

---

[*] Author to whom correspondence should be addressed.

Pathologic changes of a person's body condition are reflected in the wrist-pulse pictures. Clinical studies demonstrate that patients with hypertension, hypercholesterolemia, cardiovascular disease, and diabetes, exhibit premature loss of arterial elasticity and endothelial function, which eventually resulted in decreased flexibility of vasculature, and heightened stress to the circulatory system. The wrist-pulse shape, amplitude, and rhythm are also altered in correspondence with the hemodynamic characteristics of blood flow.

The growing recognition of the importance of developing effective preventive medical system to contemporary healthcare has placed Chinese pulse diagnosis an important position[3]. However, wrist-pulse assessment is a matter of technical skill and subjective experience. The intuitional accuracy depends upon the individual's persistent practice and quality of sensitive awareness. Different Chinese physicians might not always give identical wrist-pulse waveform pattern recognition for some given patient. The classifications of wrist-pulse waveform patterns identified and named by different Chinese physicians in their medical literatures are not always the same. In history, Chinese physicians clearly appreciated the significance of the wrist-pulses and association of changes in the wrist-pulses with diseases, but they did not progress beyond the stage of manual palpations, thereby remaining largely uninfluenced by quantitative measurements. Solid quantified description of Chinese pulse diagnosis would pave a way in its modernizing advancements.

This paper aims to extract characteristics of Chinese pulses from their waveforms by introducing quantified classification indices. These classification indices can be served for wrist-pulse waveform pattern recognition and classification in Chinese pulse diagnosis. This is a subject dealing with automated Chinese pulse diagnosis. The fact that an electronic device is interfaced with a personal computer holds open the possibility that an automated system of interpreting Chinese pulse waveform patterns could be developed.

# Methods

## Chinese pulse waveforms

Chinese pulse waveforms are recorded non-invasively with a pressure sensor. A wrist-watch-like structure is mounded to keep the sensor position well over radial artery. A sphygmomanometer cuff is wrapped around a wrist-watch-like structure to provide hold-down pressure. The Chinese pulse waveforms are captured and digitized by an analog-to-digital converter and recorded onto a personal computer. 13 Chinese pulse waveforms in TCM are recorded and their descriptions[2] are shown below in Figure 1. The 13 Chinese pulse waveforms appear more likely to help practitioner determine imbalances that relate to the selection of traditional style therapeutics, *i.e.*, acupuncture points and individual herbs.

**Figure 1** The 13 Chinese pulse waveforms and their descriptions

1. **Normal** Pulse, *Ping Mai*: A normal pulse with smooth, even, forceful, and frequency (between 60–90 beats per minute).

   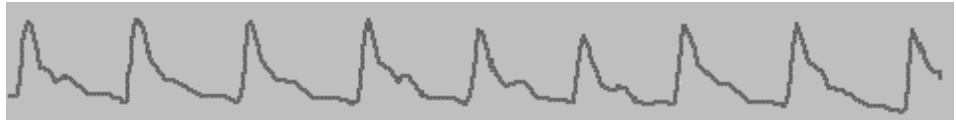

2. **Slow** Pulse, *Chi Mai*: A pulse with reduced frequency (less than 60 beats per minute), usually indicating endogenous cold.

   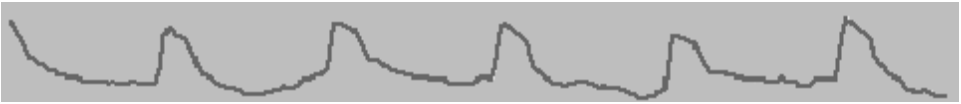

3. **Hurried** Pulse, *Cu Mai*: A rapid pulse with irregular intermittence, often due to excessive heat with stagnation of Qi and blood, or retention of phlegm or undigested food.

   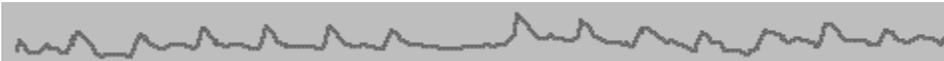

4. **Intermittent** Pulse, *Dai Mai*: A slow pulse pausing at regular intervals, often occurring in exhaustion of Zang-Fu organs, severe trauma, or being seized by terror.

   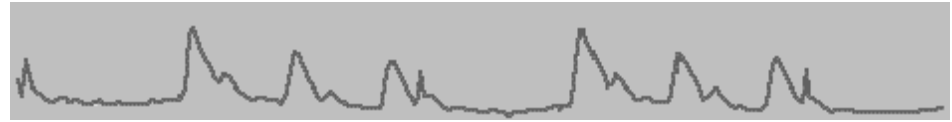

5. **Surging** Pulse, *Hong Mai*: A pulse beating like dashing waves with forceful rising and gradual decline, indicating excessive heat.

   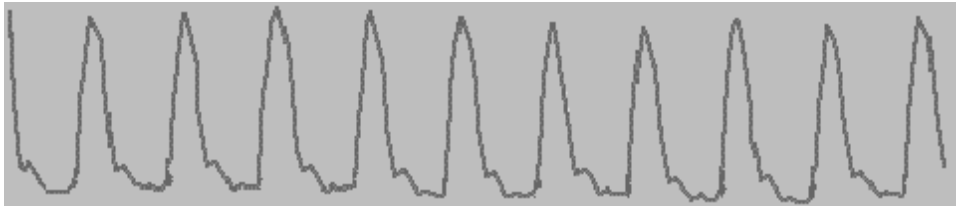

6. **Slippery** Pulse, *Hua Mai*: A pulse like beads rolling on a plate, found in patients with phlegm-damp or food stagnation, and also in normal persons. A slippery and rapid pulse may indicate pregnancy.

   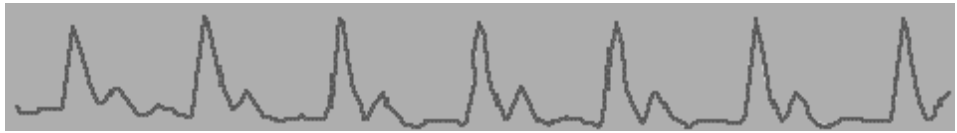

7. **Knotted** Pulse, *Jie Mai*: A slow pulse pausing at irregular intervals, often occurring in stagnation of Qi and blood.

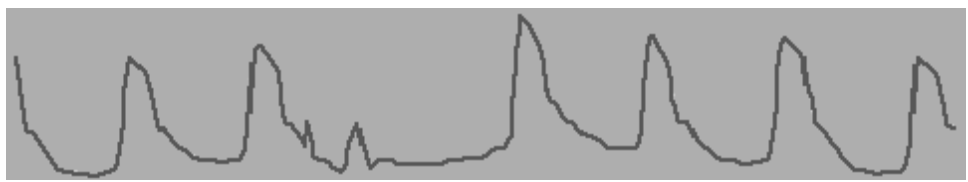

8. **Soggy** Pulse, *Ru Mai*: A superficial, thin, and soft pulse which can be felt on light touch like a thread floating on water, but grows faint on hard pressuring, indicating deficiency conditions or damp retention.

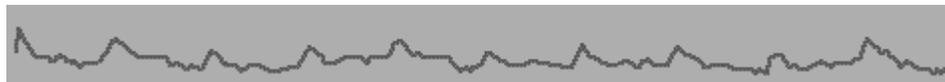

9. **Choppy** Pulse, *Se Mai*: A pulse coming and going choppily with small, fine, slow, joggling *tempo* like scraping bamboo with a knife, indicating sluggish blood circulation due to deficiency of blood or stagnation of Qi and blood.

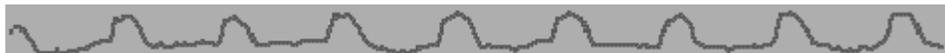

10. **Rapid** Pulse, *Shou Mai*: A pulse with increased frequency (more than 90 beats per minute), usually indicating the presence of heat.

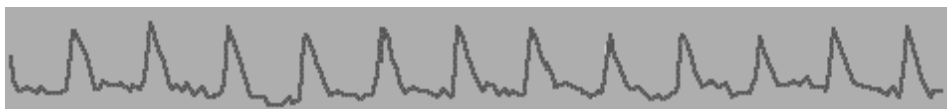

11. **Fine** Pulse, *Xi Mai*: A pulse felt like a fine thread, but always distinctly perceptible, indicating deficiency of Qi and blood or other deficiency states.

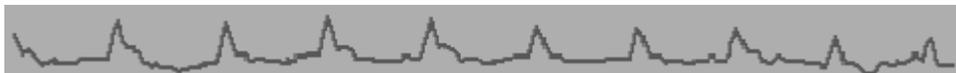

12. **Wiry** Pulse, *Xian Mai*: A pulse that feels straight and long, like a musical instrument string, usually occurring in liver and gallbladder disorders or severe pain.

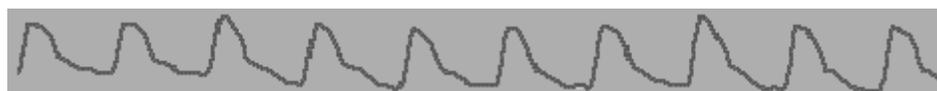

13. **Wiry & Slippery** Pulse, *Xian-Hua Mai*: A pulse with the characteristics of **Wiry** and **Slippery** Pulses existing simultaneously.

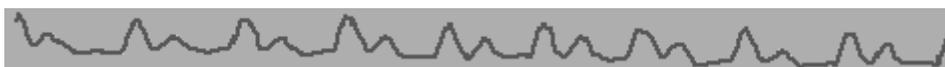

# Chinese pulse waveform analysis

A beating heart generates pressure and flow waves which propagate throughout the arterial system. The shapes of wrist-pulse waveforms are altered by their continuous interaction with the non-uniform arterial system[4-6]. The pressure waves expand the arterial walls as traveling, and the expansions are palpable as the wrist-pulses.

Each discontinuity reflects the incident waves in the mechanical and geometrical properties of the arterial tree, *e.g.*, at bifurcations and stenoses. The palpable wrist-pulses can thus be studied in terms of one forward traveling wave component, the collective waves running from heart to periphery and containing information of the heart itself; and one backward traveling wave component, the collective waves containing information of the reflection sites and the periphery of the arterial system. Moreover, the reflected pressure waves tend to increase the load to the heart and play a major role in determining the wrist-pulse waveform patterns. Hence, wrist-pulse waveforms can be expressed in terms of its forward and backward running components with a phase shift in time as illustrated in Figure 2.

**Figure 2** Forward wave is higher in amplitude, and backward wave is lower in amplitude, with a phase shift

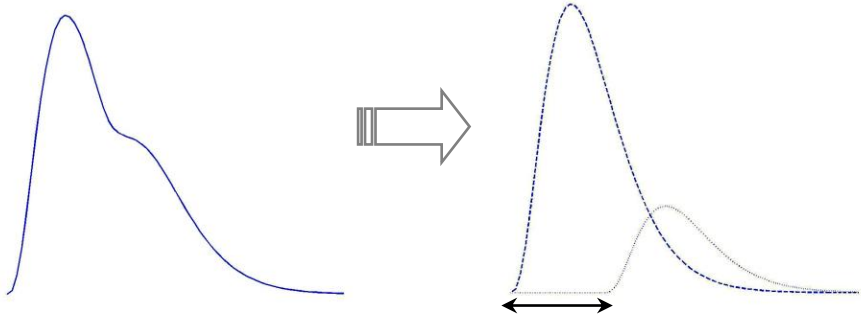

A normal wrist-pulse waveform has a smooth, fairly sharp upstroke, a momentarily sustained peak, a quick downstroke and decay. The reflected wave also has similar shape to the initial wave but smaller in amplitude[7]. Such wrist-pulse curve looks like the shape of Gamma density function. Without loss of generality, the wrist-pulse waveforms can be viewed as the summation of forward and backward waves by using the following function:

$$F(t/\alpha, \beta, \Delta, A, B) = Af(t/\alpha, \beta, 0) + Bf(t/\alpha, \beta, \Delta) = At^{\alpha} e^{-\beta t/10} + Bt^{\alpha} e^{-\beta(t-\Delta)/10},$$

where the base function is Gamma density function, formed by the product of one power function and one exponential function

$$f(t/\alpha, \beta, \Delta) = t^{\alpha} e^{-\beta(t-\Delta)/10} \qquad t \geq \Delta$$

for describing forward wave $f(t/\alpha, \beta, 0)$ and backward wave $f(t/\alpha, \beta, \Delta)$. $\alpha$ and $\beta$ are the *shape* and *rate* parameters respectively. $A$ and $B$ are amplitudes of forward and backward waves respectively. $\Delta$ is the phase shift or time delay between forward wave and backward wave. These parameters are estimated to provide the best curve fit between the 13 recorded Chinese pulse waveforms and the collective Gamma density functions.

# Results

## Chinese pulse waveform pattern recognition

In order to study the characteristic of each Chinese pulse waveform quantitatively, single period of recorded Chinese pulse has been carefully selected from each pulse wave-train. The selection is to provide typical and representative single-period Chinese pulse waveforms out of the entire wave-train for the characteristic analysis. The parameters ($\alpha, \beta, \Delta, A, B$) for each Chinese pulse are determined based on minimizing the difference between actual recorded waveform and the proposed function $F(t/\alpha, \beta, \Delta, A, B)$. In Figure 3, bubble and solid lines represent the actual recorded waveform and its corresponding fitted function $F(t/\alpha, \beta, \Delta, A, B)$ for each Chinese pulse.

**Figure 3** 13 Chinese pulses (parameters and their waveforms)

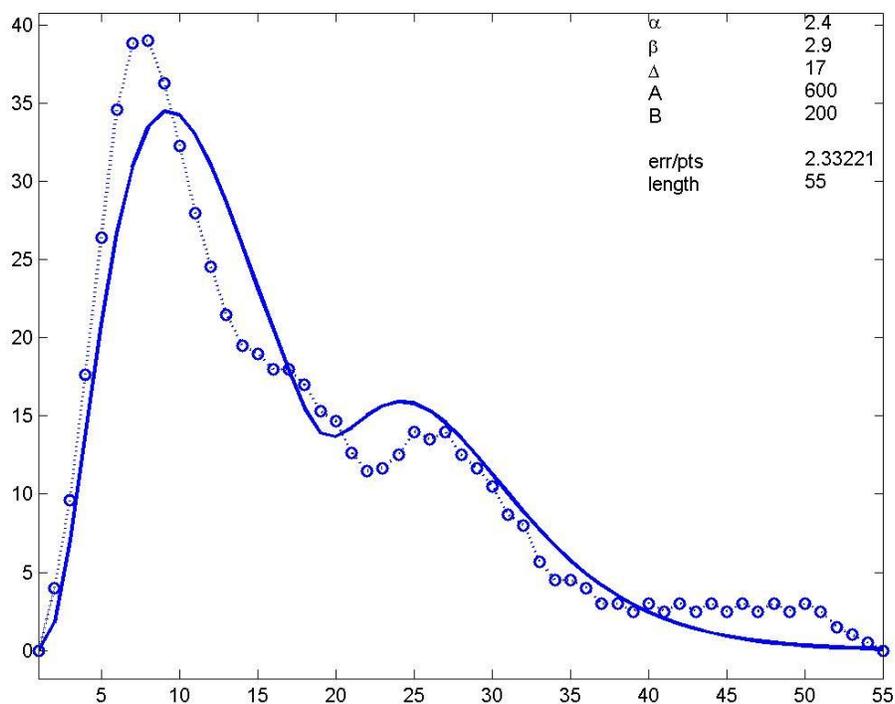

(1) **Normal** Pulse, *Ping Mai*

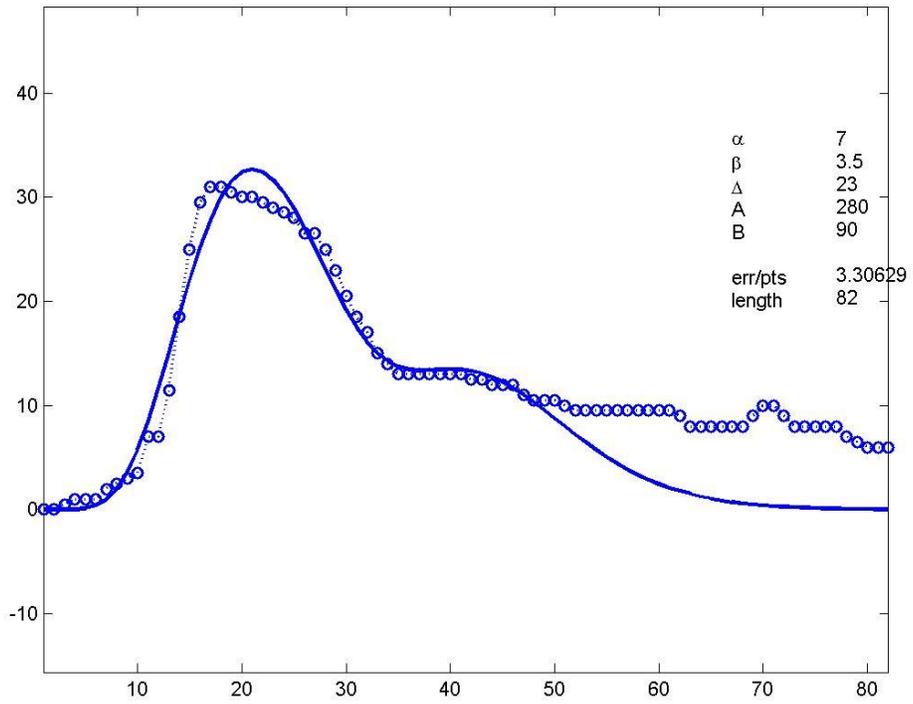

(2) **Slow** Pulse, *Chi Mai*

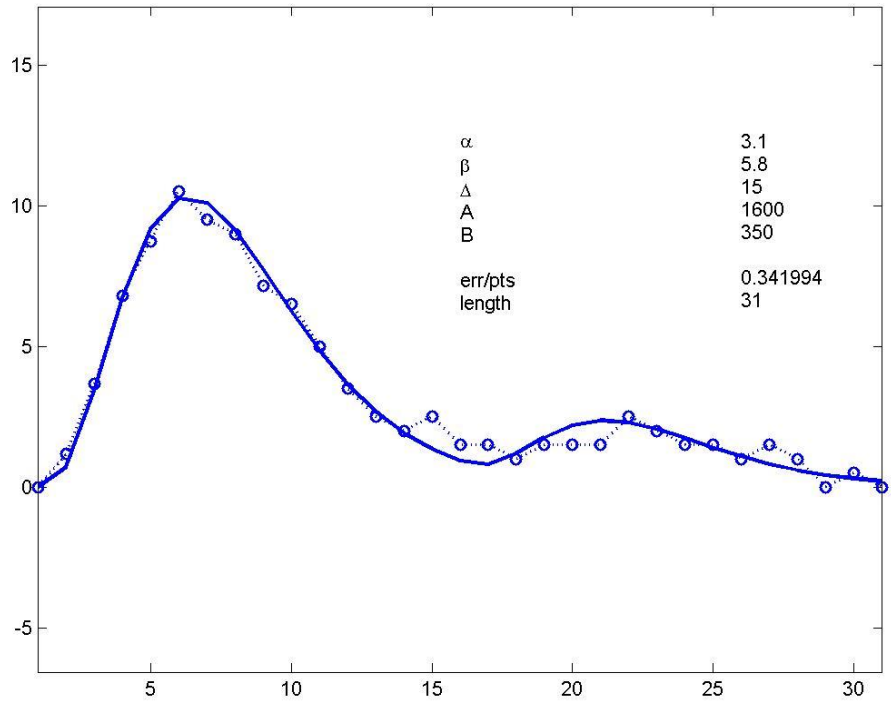

(3) **Hurried** Pulse, *Cu Mai*

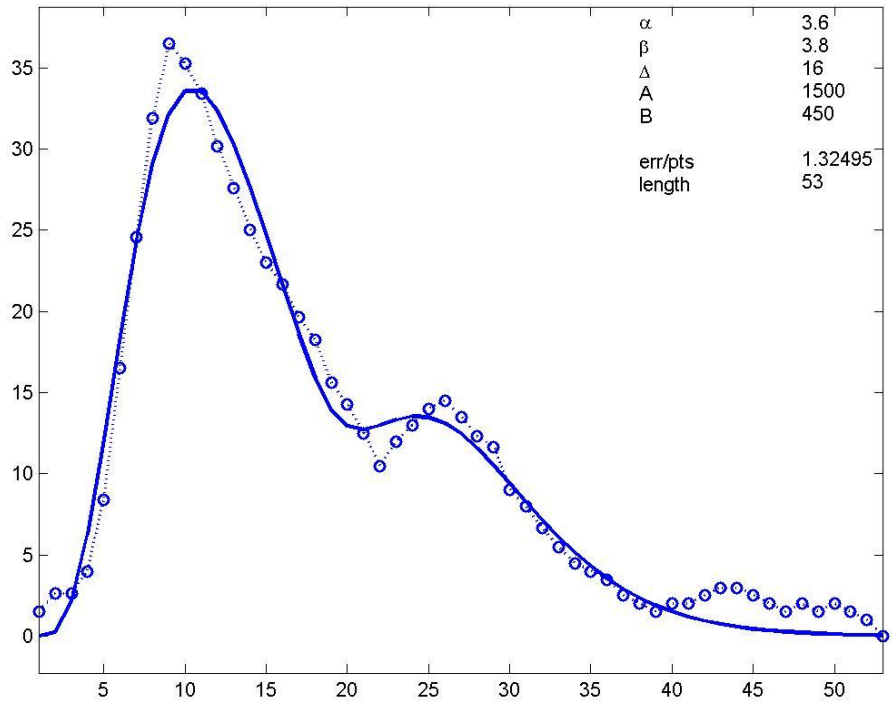

(4) **Intermittent** Pulse, *Dai Mai*

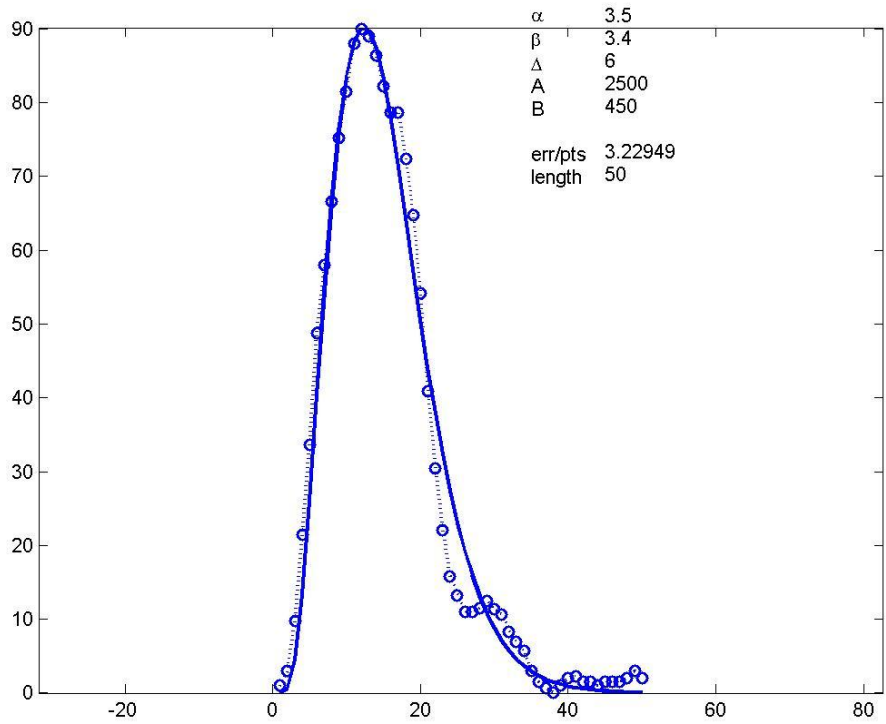

(5) **Surging** Pulse, *Hong Mai*

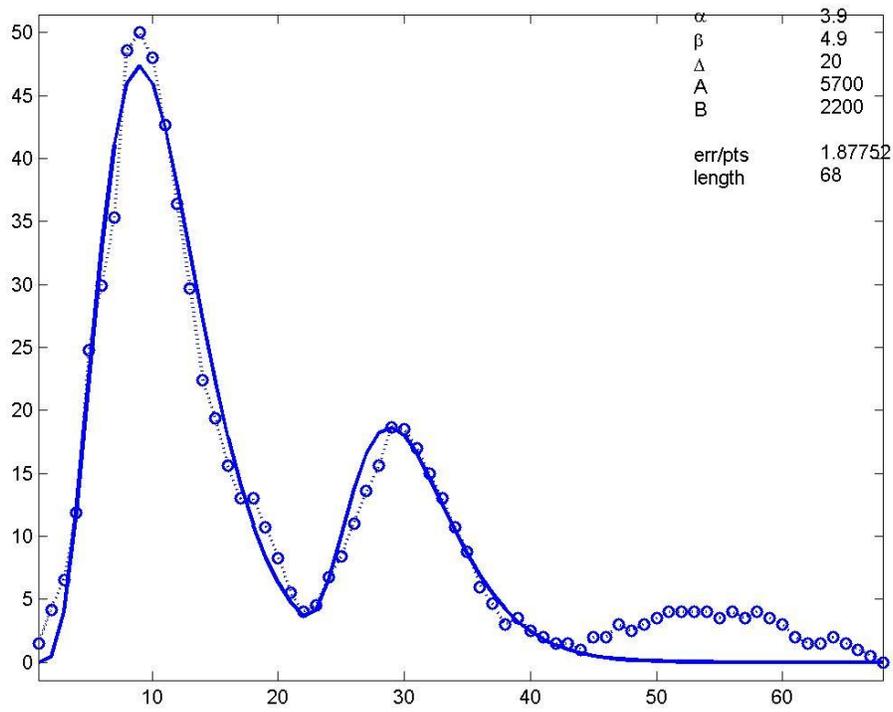

(6) **Slippery** Pulse, *Hua Mai*

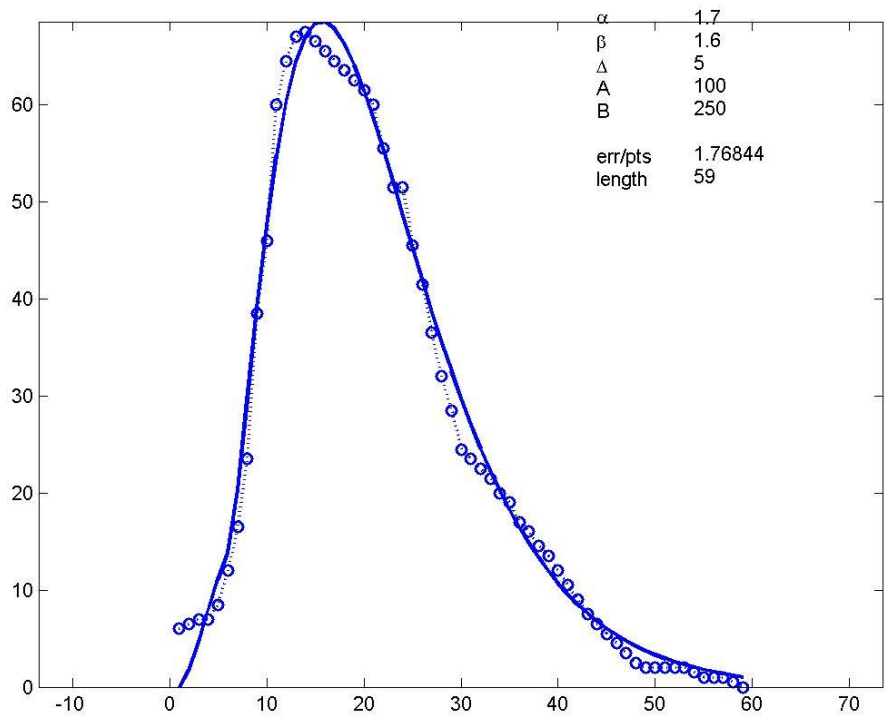

(7) **Knotted** Pulse, *Jie Mai*

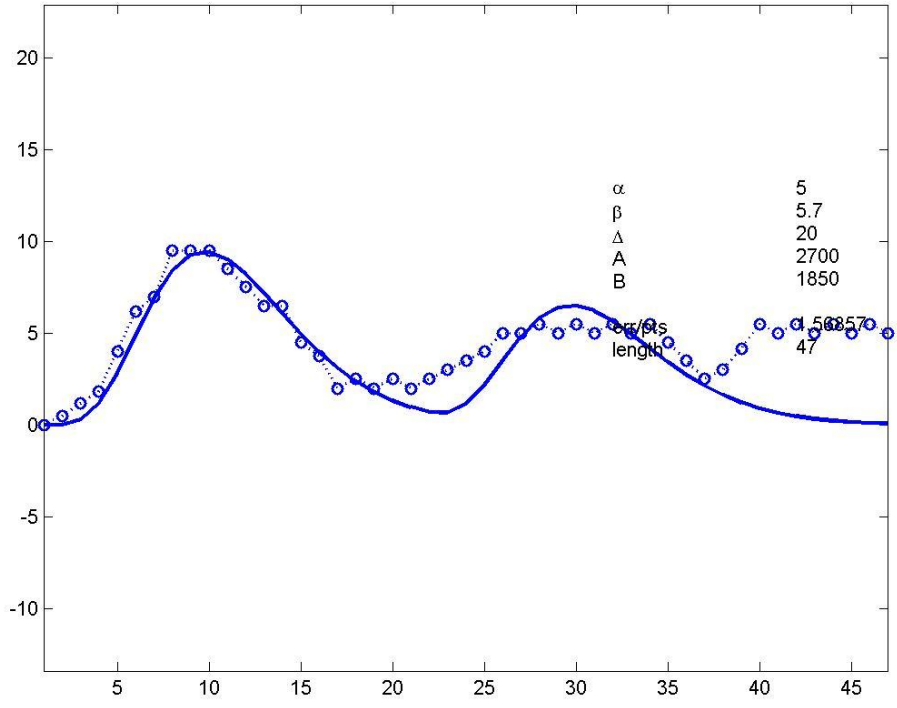

(8) **Soggy** Pulse, *Ru Mai*

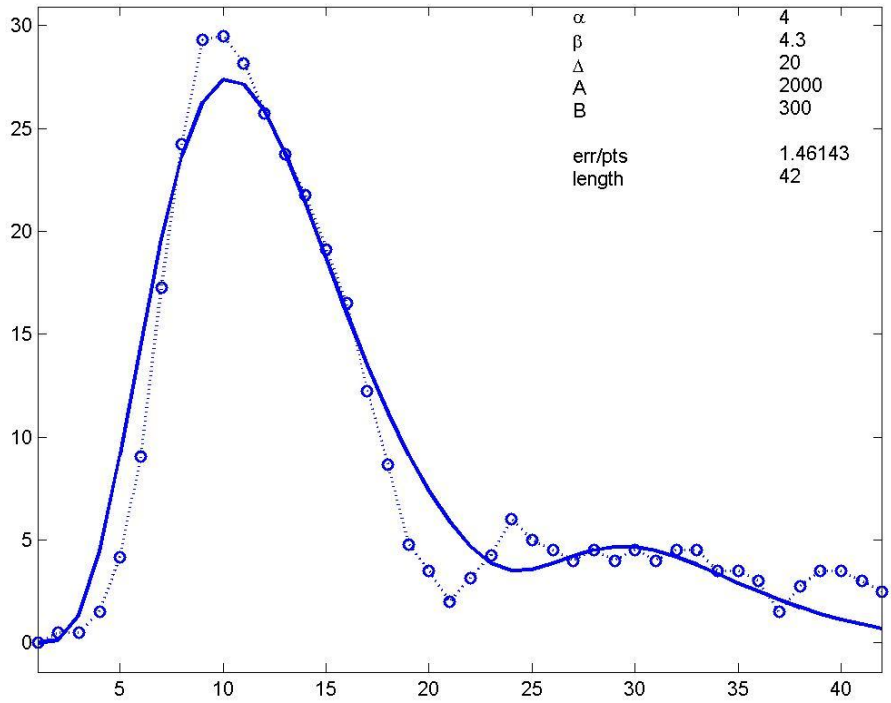

(9) **Rapid** Pulse, *Shou Mai*

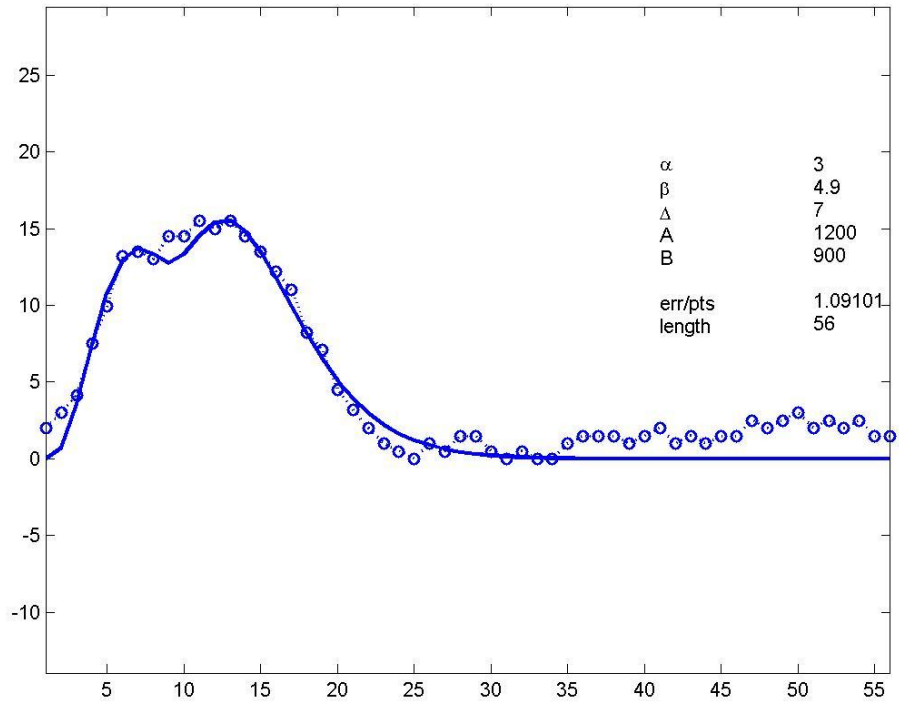

(10) **Choppy** Pulse, *Se Mai*

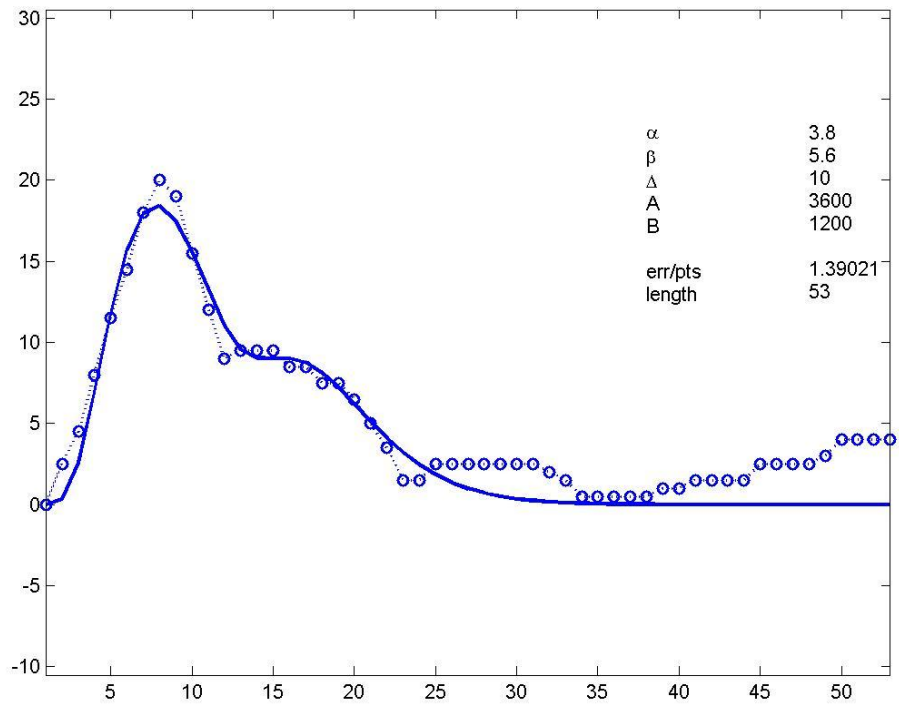

(11) **Fine** Pulse, *Xi Mai*

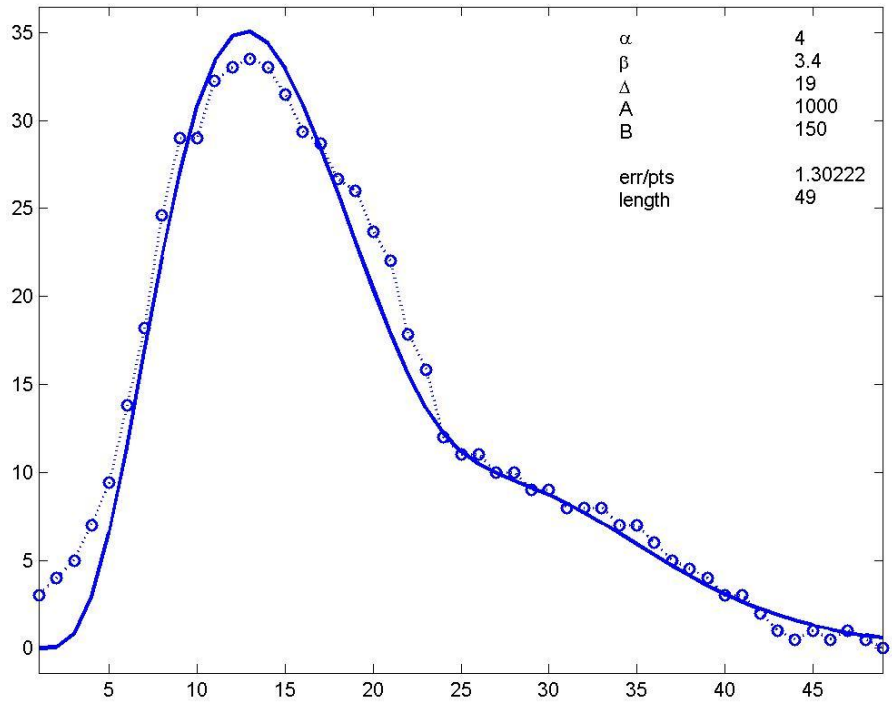

(12) **Wiry** Pulse, *Xian Mai*

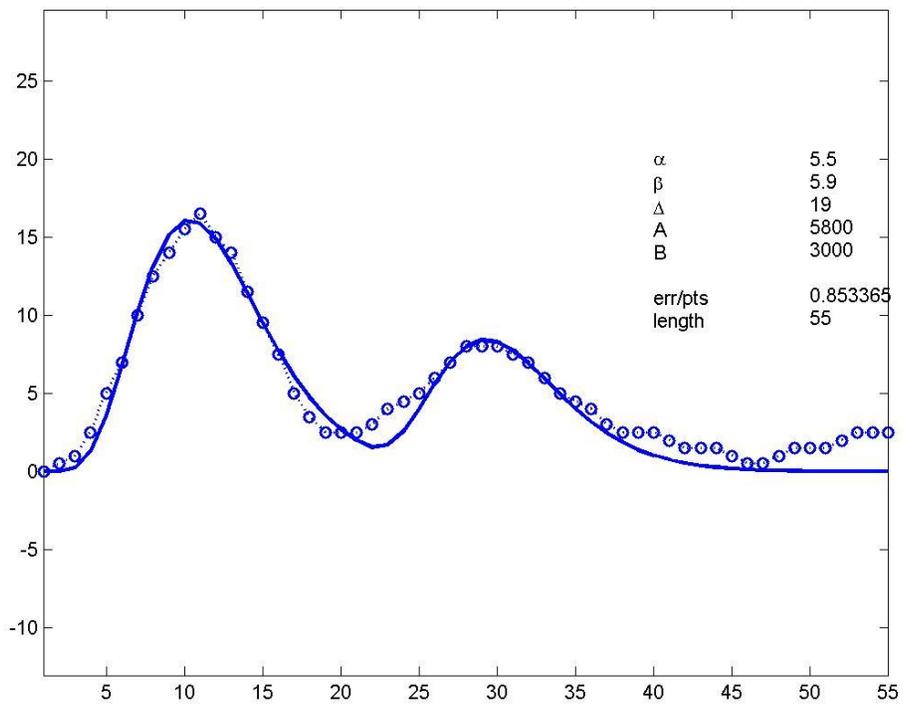

(13) **Wiry & Slippery** Pulse, *Xian-Hua Mai*

The values of parameters ($\alpha, \beta, \Delta, A, B$) in the model for each Chinese pulse are summarized and listed in Table 1 below. The parameters in the Table 1 have been arranged according to the wave length ($L$). The parameters $L$, $10L\beta/\alpha$, $A/B$ and $\Delta/L$ are also shown in the Table 1.

**Table 1** Classification indices

| Pulses | $\alpha$ | $\beta$ | $\Delta$ | $A$ | $B$ | $L$ | $10L\beta/\alpha$ | $A/B$ | $\Delta/L$ |
|---|---|---|---|---|---|---|---|---|---|
| **Hurried** | 3.1 | 5.8 | 15 | 1600 | 350 | 31 | 0.60 | 4.57 | 0.48 |
| **Rapid** | 4 | 4.3 | 20 | 2000 | 300 | 42 | 0.26 | 6.67 | 0.48 |
| **Soggy** | 5 | 5.7 | 20 | 2700 | 1850 | 47 | 0.24 | 1.46 | 0.43 |
| **Wiry** | 4 | 3.4 | 19 | 1000 | 150 | 49 | 0.17 | 6.67 | 0.39 |
| **Surging** | 3.5 | 3.4 | 6 | 2500 | 450 | 50 | 0.19 | 5.56 | 0.12 |
| **Intermittent** | 3.6 | 3.8 | 16 | 1500 | 450 | 53 | 0.20 | 3.33 | 0.30 |
| **Fine** | 3.8 | 5.6 | 10 | 3600 | 1200 | 53 | 0.28 | 3.00 | 0.19 |
| **Normal** | 2.4 | 2.9 | 17 | 600 | 200 | 55 | 0.22 | 3.00 | 0.31 |
| **Wiry & Slippery** | 5.5 | 5.9 | 19 | 5800 | 3000 | 55 | 0.20 | 1.93 | 0.35 |
| **Choppy** | 3 | 4.9 | 7 | 1200 | 900 | 56 | 0.29 | 1.33 | 0.13 |
| **Knotted** | 1.7 | 1.6 | 5 | 100 | 250 | 59 | 0.16 | 0.40 | 0.08 |
| **Slippery** | 3.9 | 4.9 | 20 | 5700 | 2200 | 68 | 0.18 | 2.59 | 0.29 |
| **Slow** | 7 | 3.5 | 23 | 280 | 90 | 82 | 0.06 | 3.11 | 0.28 |

$L$ is the length of wave. $10L\beta/\alpha$ gives the point where the wave reaches its peak value. $A/B$ and $\Delta/L$ are the peak ratio and the relative phase difference between forward and backward waves respectively.

## Classification of Chinese pulses

The values of the parameters describe characteristic of Chinese pulses in recognizing their waveform patterns.

I. Whether Chinese pulse is rapid or slow is an important quality to be determined in Chinese pulse diagnosis. The length ($L$) of the Chinese pulse wave, which reflects its period or frequency, can firstly be considered to classify the 13 Chinese pulse waveforms. As shown in the results, **Hurried** pulse is very small in length ($L$), while **Slow** pulse is large. These two Chinese pulses stand out of the other 11 Chinese pulses in their length ($L$) and can be recognized by the length easily. The rest 11 Chinese pulses are undistinguishable purely by examining their lengths at this stage and should be grouped together.

II. Next, $\Delta/L$ should be used to classify the rest 11 Chinese pulses. The value $\Delta/L$ shows the relative phase difference between the forward and backward waves in each Chinese pulse waveform. The 11 Chinese pulses can be classified into 4 groups by $\Delta/L$. They are {**Knotted**, **Surging**, **Choppy**}, {**Fine**}, {**Slippery**, **Intermittent**, **Normal**, **Wiry & Slippery**, **Wiry**} and {**Soggy**, **Rapid**}. At this stage, **Fine** Pulse is obviously recognizable, while other Chinese pulses are undistinguishable in their respective groups.

III. $\beta$ should be used to classify the rest. At this stage, **Knotted**, **Surging**, **Choppy**, **Normal**, **Slippery**, **Wiry & Slippery**, **Rapid** and **Soggy** pulse are recognizable by examining $\beta$. Only **Intermittent** pulse and **Wiry** pulse are left undistinguishable, remaining in group.

IV. Finally these two Chinese pulses are recognizable by examining $A/B$.

The procedure of Chinese pulse grouping and classification is illustrated in Table 2 shown below.

**Table 2** Chinese wrist-pulse distinguishing and grouping

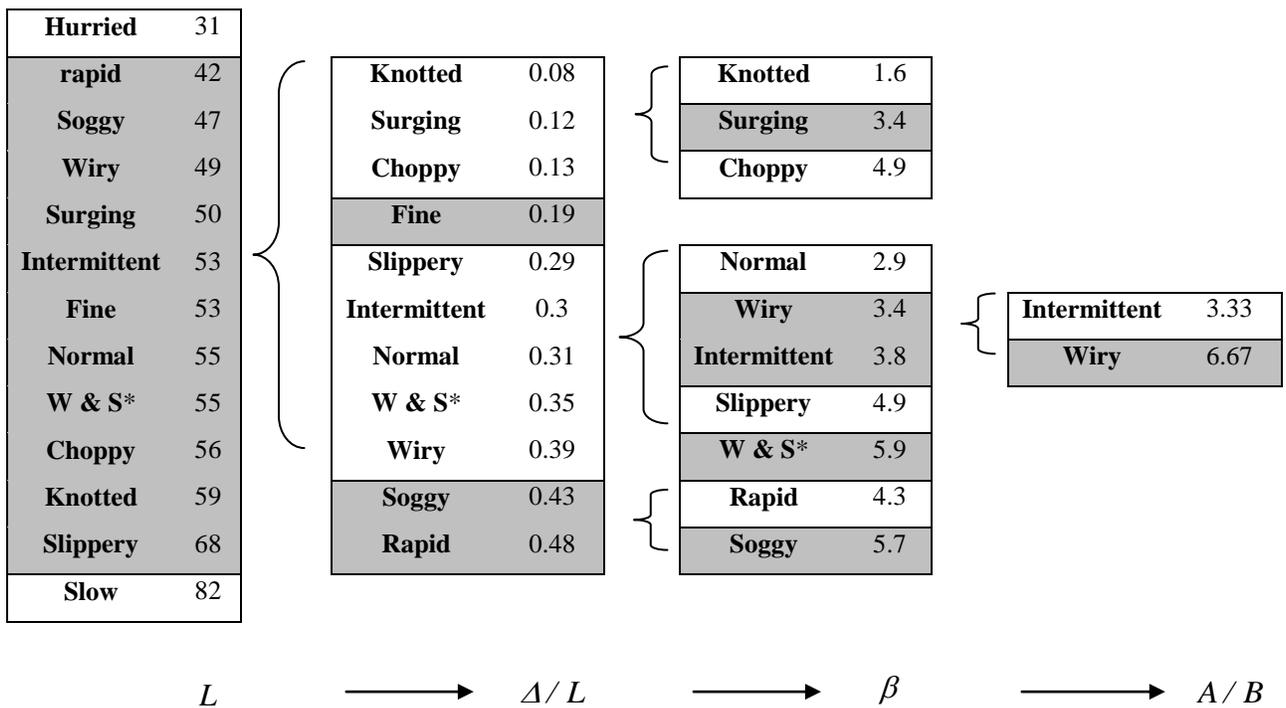

*$W \& S$ is short for **Wiry & Slippery**

According to the grouping and classification above, the ranges of parameters to group and classify the Chinese pulses in each stage are set in Table 3.

**Table 3** Chinese pulse pattern recognition process

| Cu | Shou | Ru | Xian | Hong | Dai | Xi | Ping | Xian Hua | Se | Jie | Hua | Chi |
|---|---|---|---|---|---|---|---|---|---|---|---|---|
| Hurried | Rapid | Soggy | Wiry | Surging | Intermittent | Fine | Normal | Wiry & Slippery | Choppy | Knotted | Slippery | Slow |
| $L$=24-39 | | | | | $L$=40-70 | | | | | | | $L$=71-85 |

| | Knotted Surging Choppy | Fine | Slippery Intermittent Normal | Wiry & Slippery | Wiry | Soggy Rapid | |
|---|---|---|---|---|---|---|---|
| | $\Delta/L$=0–0.16 | $\Delta/L$=0.17-0.24 | $\Delta/L$=0.25–0.41 | | | $\Delta/L$=0.42–0.60 | |

| | Knotted | Surging | Choppy | | Normal | Wiry | Intermittent | Slippery | Wiry & Slippery | Soggy | Rapid | |
|---|---|---|---|---|---|---|---|---|---|---|---|---|
| | $\beta$=1.3–2.2 | $\beta$=3.3–4.2 | $\beta$=4.3–5.2 | | $\beta$=2.3-3.2 | $\beta$=3.3–4.2 | | $\beta$=4.3–5.2 | $\beta$=5.3-7.2 | $\beta$=4.3-5.2 | $\beta$=5.3-7.2 | |

| Intermittent | Wiry |
|---|---|
| $A/B$=2.0-4.9 | $A/B$=5.0-9.0 |

# Discussion

Chinese pulse diagnosis plays an important role in patient evaluation in Traditional Chinese Medicine. Pathologic changes of a person's body condition are reflected in wrist-pulse waveforms. Chinese physicians use fingertips to diagnose the wrist-pulses of patients in order to determine their health conditions. Hence different Chinese physicians may interpret a pathological wrist-pulse waveform in a totally different way. The fact that an electronic device is interfaced with a personal computer holds open the possibility that an automated system of interpreting Chinese pulse waveform patterns relating to the health conditions of patients could be developed. In order to perform the automated Chinese pulse diagnosis, the following new classification indices have been introduced in this paper to extract characteristics of Chinese pulses from their waveforms.

- Rate parameter $\beta$ -- governing the shape of waveform and the property of the waveform decay,
- Phase difference $\Delta$ -- phase difference between forward wave and backward wave,
- Relative phase difference $\Delta/L$ -- phase difference relative to the entire period,
- Wave length $L$ -- period of wave,
- Peak ratio $A/B$ -- ratio of forward wave peak to backward wave peak.

Chinese pulses can be recognized quantitatively by the newly-developed classification indices ($L$, $\Delta/L$, $\beta$, $A/B$). The new quantitative classification not only reduces the dependency of pulse diagnosis on Chinese physician's experience, but also is able to interpret pathological wrist-pulse waveforms more precisely. The quantitative information will help us better to characterize or differentiate the wrist-pulse waveforms, independently to confirm the health states of patients, and thereby to develop means for non-invasively diagnosing a myriad of diseases associated with malfunctioning organ systems.